\documentclass[preprint,prd,amsmath,amssymb,amsthm,nofootinbib]{revtex4}
\usepackage[mathscr]{euscript}
\usepackage{bm}% bold math
\usepackage{textcomp}
\usepackage{graphicx}% Include figure files
\usepackage{subfigure}
\usepackage{overpic}
\usepackage{multirow}
\usepackage{floatrow}
\usepackage{float} %for fixing the figure
\usepackage{amsmath,amssymb,amsthm}
\usepackage{cases}
\usepackage[colorlinks=true,linkcolor=red]{hyperref}
\newfloatcommand{capbtabbox}{table}[][\FBwidth]
\newcommand{\bea}{\begin{eqnarray}}
\newcommand{\eea}{\end{eqnarray}}
\newcommand{\beq}{\begin{equation}}
\newcommand{\eeq}{\end{equation}}

\def\/{\over}

\begin{document}

\title{Resonant amplification of curvature perturbations in inflation model with periodical derivative coupling}

\author{ Li-Yang Chen$^{1}$\footnote{clyrion@hunnu.edu.cn}, Hongwei Yu$^{1,2}$\footnote{hwyu@hunnu.edu.cn} and Puxun Wu$^{1,2}$\footnote{pxwu@hunnu.edu.cn}  }
\affiliation{$^{1}$Department of Physics and Synergetic Innovation Center for Quantum Effects and Applications, Hunan Normal University, Changsha, Hunan 410081, China\\
$^{2}$Institute of Interdisciplinary Studies, Hunan Normal University, Changsha, Hunan 410081, China
}

\begin{abstract}
 In this paper, we introduce  a weak,  transient and periodical derivative  coupling between   the inflaton field and gravity, and  find that the  square of the sound speed of the curvature perturbations becomes a periodic function, which results in that the equation of the curvature perturbations can be transformed into the form of   the Mathieu equation  in the sub-horizon limit.  Thus, the parametric resonance will amplify  the curvature perturbations so as to generate  a formation of   abundant primordial black holes (PBHs). We show that the generated PBHs can make up  most of  dark matter.  Associated  with the generation of PBHs, the large scalar perturbations will give rise to the scalar induced gravitational waves which may be  detected by future gravitational wave projects.  \end{abstract}

%\pacs{98.80.Cq, 04.50.Kd, 05.70.Fh}

\maketitle
%%%%%%%%%%%%%%%%%%%%%%%%%%%%%%%%%%%%%%%%%%%%%%%%%%
\section{Introduction}
\label{sec_in}

When the large curvature perturbations produced during  inflation re-enter the Hubble horizon in the radiation- or matter-dominated epoch, the gravitational collapse of  dense regions  will result in  generation of primordial black holes (PBHs)~\cite{Zeldovich, Hawking, Carr, Meszaros, Carr1975, Khlopov, Ozsoy2023, Choudhury2023f}. The PBHs have been used to explain the gravitational wave (GW) events generated by  merger of massive black holes observed by the LIGO/Virgo collaboration~\cite{lg1,lg2,lg3,lg4}, the six ultrashort-timescale microlensing events in the OGLE data~\cite{P.Mroz2017,H.Niikura2019}, and the ninth asteroid in the solar system  \cite{J.Scholtz2020,E.Witten}. Furthermore, the asteroid-mass PBHs can make up all dark matter~\cite{A.Katz2018,H.Niikura2019a,A.Barnacka2012,P.W.Graham2015}. Accompanying the formation of  PBHs, the large scalar perturbations will become an important GW source and lead to the so-called scalar induced GWs (SIGWs). The SIGWs may be detected by the future GW projects such as LISA \cite{lisa} Taiji ~\cite{taiji}, TianQin~\cite{tianqin} and PTA ~\cite{pta1,pta2,pta3,pta4}.   An observation of such SIGWs would provide evidence for the existence of PBHs. 

However, in the standard slow-roll inflation, which predicts  nearly scale-invariant curvature perturbations, the possibility of the formation of PBHs is negligible. This is because the cosmic microwave background radiation  (CMB) observations have implied a very small  amplitude of the power spectrum of the curvature perturbations of about $\mathcal{O}(10^{-9})$. To generate a sizable amount of PBHs requires that the amplitude of the power spectrum of the curvature perturbations  reaches about $\mathcal{O}(10^{-2})$.  
Since the CMB observations only give a limit on the curvature perturbations at large scales~\cite{Aghanim2020} and  the magnitude of the power spectrum at scales smaller than the CMB one is not restricted strongly by any observations,  the formation of an abundant PBHs will be possible if there are some mechanisms to enhance the curvature perturbations at small scales.

  It is well known that  the amplitude of the power spectrum of the curvature perturbations $\mathcal{R}$  is given by $\mathcal{P_R}=\frac{H^2}{8\pi^2 \epsilon c_s}$ when the mode exits the horizon during inflation in the standard slow-roll inflation. Here $\epsilon$ is the slow-roll parameter, which is proportional to the rolling speed of the inflaton, $c_s$ is the sound speed of the curvature perturbations and  $H$ is the Hubble parameter. So,  a natural way to enhance the curvature perturbations is to reduce the rolling speed of the inflaton~\cite{Yokoyama1998,Choudhury2014,Germani2017,Motohashi2017,H. Di2018, Ballesteros2018, Dalianis2019,Gao2018,Tada2019, Mishra2020, Atal2020,Ragavendra2021, Bhaumik2020,   Drees2021,C.Fu2020, Xu2020, Lin2020, Dalianis2021, Yi2021,Gao2021, Yi2021b, TGao2021, Solbi2021,Gao2021b, Solbi2021b,  Zheng2021,Teimoori2021a, Cai2021, Wang2021, Fuchengjie2019,fuchengjie2020,Dalianis2020,Teimoori2021,Karam2022, Heydari2022, Heydari2022b,Bellido2017,Ezquiaga2018,Pi2022, Choudhury2023cc,Meng2022,Mu2022,Kawaguchi2022,Fu2022,Chen2022,Gu2023,Yi2022,S.Choudhury2023,Dalianis2023,Zhao2023,Heydari2023,Choudhury2023a,Firouzjahi2023,Mishra2023,Asadi2023,Cole2023,Choudhury2023b,Frolovsky2023,Chen2023a,Su2023,Cable2023,Escriva2023,Choudhury2023c,Choudhury2023e,Choudhury2024} or to suppress the sound speed $c_s$~\cite{G. Ballesteros,Kamenshchik,Gorji2022,Romano3,Ballesteros2022,R. Zhai,Qiu2022,Zhai2023}.
   Additionally, some other ways can also predict the formation of PBHs \cite{Choudhury2023d,Corral2023,Kasai2023,Stamou2023,Ge2023,Bhattacharya2023,Ijaz2023,Cheung2023,Mansoori2023,Gow2023,Li2023,Tada2023,Cai2023y,Huang2023a,Ghoshal2023,Saburov2023,Tada2023a,Huang2023,Arya2023,Cai2023i,Martin2020a,Martin2020b}.  
 The decrease of the rolling speed of the inflaton can be realized by flattening the potential of the inflaton field. The corresponding inflation model is  called  the inflection-point inflation~\cite{Germani2017,Motohashi2017,Ezquiaga2018,H. Di2018,Ballesteros2018,Dalianis2019,Gao2018,Drees2021,C.Fu2020, Xu2020, Lin2020, Dalianis2021, Yi2021,Gao2021, Yi2021b, TGao2021, Solbi2021, Gao2021b, Solbi2021b, Zheng2021,Teimoori2021a, Cai2021, Wang2021}. Recently, it has been shown that the gravitationally enhanced friction  can also slow the rolling speed of the inflaton~\cite{Fuchengjie2019,fuchengjie2020,Dalianis2020,Teimoori2021, Heydari2022, Heydari2022b}.  In this mechanism, a derivative coupling between the inflaton field and  gravity is invoked. In addition, the growth of curvature perturbations caused by parametric resonance has also been extensively studied~\cite{yfcai2018,yfcai2019,c.chen2019,c.chen2020,Addazi2022,Cai2020,bLi2023}. The parametric resonance can be obtained by adding a periodic correction to the potential of the inflaton field or considering a periodic sound speed.  
As the derivative coupling can realize the decrease of the inflaton's rolling speed, can it lead to the parametric resonance to amplify the curvature perturbations? This motivates us to finish the present study. We find that the parametric resonance will occur after introducing a periodical derivative coupling between the inflaton field and gravity,
since  the equation of the curvature perturbations can be transformed into the form of the Mathieu equation.   We demonstrate that  
 the enhanced curvature perturbations will lead to an abundant generation of PBHs, which can make up most of  dark matter, 
 and the SIGWs may be detected by the future GW projects.

The rest of this paper is organized  as follows: In Sec.~\ref{sec2}, we will introduce the inflation model with a periodical  derivative coupling between  the inflaton field and  gravity. Sec.~\ref{sec3} discusses the parametric resonance and Sec.~\ref{sec4} describes the formation of PBHs. In Sec.~\ref{sec5}, we investigate  the  SIGWs. Finally, we give our conclusions in Sec.~\ref{conclusion}.

\section{inflation with periodical derivative coupling}
\label{sec2}

We consider an inflation model with a non-minimal derivative coupling between the  inflaton field $\phi$  and  gravity. The action of the system has the form
 \begin{align}\label{action}
	\mathcal{S}=\int d^{4} x \sqrt{-g}\left[\frac{M_{\mathrm{pl}}^2}{2} R-\frac{1}{2}\left(g^{\mu \nu}-\frac{1}{M_{\mathrm{pl}}^2}  \theta(\phi) G^{\mu \nu}\right) \nabla_{\mu} \phi \nabla_{\nu} \phi-V(\phi)\right].
 \end{align}
Here $g$ is the determinant of the metric tensor $g_{\mu\nu}$,  $M_{\mathrm{pl}} $ is the reduced Planck mass, $R$ is  the Ricci scalar,  $G_{\mu\nu}$ is the Einstein tensor, $\theta(\phi)$ denotes  the coupling function, and $V(\phi)$ is the potential of the inflaton field.  This action  belongs to a class of the general Horndeski's theories with second-order equations of motion~\cite{Deffayet2011,Kobayashi2011}, which can be free of the ghost and gradient instabilities~\cite{Kobayashi2011}. The Lagrangian of such Horndeskis theories has the term $G_5(\phi,X) G_{\mu\nu}\nabla_\mu \nabla_\nu \phi$, where $G_5$ is a generic function of $\phi$ and $X \equiv -\partial_\mu\phi \partial^\nu \phi/2$. By choosing the function $G_5=-\kappa^2 \chi(\phi)/2$, the term containing $\theta$  in Eq.~(\ref{action}) can be recovered from the Horndeski's Lagrangian after integration by parts with $\theta$ being defined as $\theta\equiv d\chi/d\phi$.

In the spatially flat Friedmann-Robertson-Walker background, the background equations derived  from the action (\ref{action}) are   
\begin{align}\label{br1} 
	 3 H^{2}=\frac{1}{M_{\mathrm{pl}}^{2}}\left[\frac{1}{2}\left(1+ \frac{9}{M_{\mathrm{pl}}^{2}} \theta(\phi) H^{2}  \right) \dot{\phi}^{2}+V(\phi)\right],
\end{align}
\begin{align}\label{br2} 
	-2 \dot{H}=\frac{1}{M_{\mathrm{pl}}^{2}}\left[\left(1+\frac{3}{M_{\mathrm{pl}}^{2}} \theta(\phi) H^{2}-\frac{1}{M_{\mathrm{pl}}^{2}} \theta(\phi) \dot{H}\right) \dot{\phi}^{2}- \frac{1}{M_{\mathrm{pl}}^{2}} \theta_{,\phi}(\phi) H \dot{\phi}^{3} - \frac{2}{M_{\mathrm{pl}}^{2}}  \theta(\phi) H \dot{\phi} \ddot{\phi}\right],
\end{align}
\begin{eqnarray}\label{br3} 
	\left(1+ \frac{3}{M_{\mathrm{pl}}^{2}} \theta(\phi) H^{2}\right) \ddot{\phi}+\left[1+\frac{1}{M_{\mathrm{pl}}^{2}} \theta(\phi)\left(2 \dot{H}+3 H^{2}\right)\right] 3 H \dot{\phi} +  \frac{3}{2M_{\mathrm{pl}}^{2}} \theta_{,\phi}(\phi) H^{2} \dot{\phi}^{2}+V_{,\phi}(\phi)=0.
\end{eqnarray}
 Here  the overdot denotes the derivative with respect to the cosmic time $t$,  $\theta_{,\phi}\equiv d\theta/d\phi$ and $V_{,\phi}\equiv dV/d\phi$. 
 
To describe the slow-roll inflation, 
 we define the slow-roll parameters 
\begin{align}\label{SLC}
\epsilon &=-\frac{\dot{H}}{H^{2}}, \quad \delta_{\phi} =\frac{\ddot{\phi}}{H \dot{\phi}}, \nonumber\\
\delta_{X} &=\frac{ \dot{\phi}^{2}}{2  {M_{\mathrm{pl}}^{2}}H^{2}}, \quad \delta_{D} =\frac{  \theta \left(\phi \right) \dot{\phi}^{2}}{4{M_{\mathrm{pl}}^{4}}}.
\end{align}
When $\{\epsilon, |\delta_{\phi}|, \delta_{X}, |\delta_{D}| \}\ll 1$ are satisfied,  the slow-roll inflation  is obtained. Furthermore, we add the condition: $|\frac{3\theta(\phi) H^2}{M_\mathrm{pl}^2}|\ll1$. %  and $|\frac{\dot{\theta}(\phi) H}{M_\mathrm{pl}^2}| \ll1$. 
Thus, the background dynamics in the non-minimally derivative coupled inflation model  will be almost the same as that in the minimal coupling case

During inflation,  the quantum fluctuations   provide the seed for the formation of  large scale cosmic structures. The fluctuations are described by using  the curvature perturbations $\mathcal{R}$.  Expanding the action given in Eq.~(\ref{action}) to the second-order, one can obtain the action of $\mathcal{R}$~\cite{A.D.Felice2011, Germani2010, S.Tsujikawa2012,Kobayashi2011} 
\begin{align}\label{S2}
S^{(2)}=\int d t d^{3} x a^{3} Q\left[\dot{\mathcal{R}}^{2}-\frac{c_{s}^{2}}{a^{2}}(\partial \mathcal{R})^{2}\right],
\end{align}
where
\bea\label{Q}
Q=\frac{w_{1}\left(4 w_{1} w_{3}+9 w_{2}^{2}\right)}{3 w_{2}^{2}},
\eea
and 
\bea
c_{s}^{2}=\frac{3\left(2 w_{1}^{2} w_{2} H-w_{2}^{2} w_{4}+4 w_{1} \dot{w}_{1} w_{2}-2 w_{1}^{2} \dot{w}_{2}\right)}{w_{1}\left(4 w_{1} w_{3}+9 w_{2}^{2}\right)}
\eea
with 
\begin{align} \label{wi}
w_{1}&=M_{\mathrm{pl}}^{2}\left(1-2 \delta_{D}\right) , \nonumber \\
w_{2}&=2 H M_{\mathrm{pl}}^{2}\left(1-6 \delta_{D}\right), \nonumber \\
w_{3}&=-3 H^{2} M_{\mathrm{pl}}^{2}\left(3-\delta_{X}-36 \delta_{D}\right), \nonumber \\
w_{4}&=M_{\mathrm{pl}}^{2}\left(1+2 \delta_{D}\right).
\end{align}
 Defining $z \equiv  a\sqrt{2Q}$ and $u_{k} \equiv  z\mathcal{R}_{k}$, we find that $u_k$ satisfies the equation 
  \begin{align}\label{sasaki2}
\ddot{u}_{k}+H\dot{u}_{k}+\left( \frac{c_{s}^{2}k^{2}}{a^{2}} -\frac{\ddot{z}+H\dot{z}}{z}\right) u_{k}=0. 
\end{align}
Here $k$ is the wave-number and $a$ is the cosmic scale factor. Solving Eq.~(\ref{sasaki2}) leads to the power spectrum of the curvature perturbations 
\begin{align}\label{prk}
\mathcal{P}_\mathcal{R}\left(k\right)=\frac{k^{3}}{2 \pi^{2}} \left|\frac{u_{k}}{z}\right|^{2}.
\end{align}
The CMB observations have implied that at large scales this power spectrum is  a nearly scale-invariant spectrum with the amplitude being about $\mathcal{O}(10^{-9})$~\cite{Aghanim2020}.

To generate a sizable amount of PBHs, we need to enhance the curvature perturbations at scales smaller than the CMB one through the  parametric resonance. Thus, we choose  the coupling function $\theta(\phi)$  to take an oscillating  form
\begin{align}
\theta(\phi)=w \sin \left(\frac{\phi}{\phi_{c}}\right) \Theta\left(\phi_{s}-\phi\right) \Theta\left(\phi-\phi_{e}\right).
\end{align}
Here $w$ is a dimensionless constant, which must satisfy  $\omega \ll \frac{M_\mathrm{pl}^2}{3H^{2}}$ since $|\frac{3\theta(\phi)H^2}{M_\mathrm{pl}^2}| \ll1$, and $\phi_{c}$ is a quantity with the same dimension as $\phi$ and is set to be much less than $\phi$. Meanwhile $\phi_{s}$ and $\phi_{e}$ represent the beginning and the end of the coupling, respectively, and  $\Theta$  is the unit Heaviside step function.  The value of $\phi_s$ is chosen to be away from that of $\phi$ at the beginning of inflation, and thus the derivative coupling does not affect the curvature perturbations at the CMB scale. So, the amplitude of the power spectrum of the curvature perturbations at the CMB scale remains to be of the standard form:   $ \mathcal{P}_{\mathcal{R}_0}= \frac{V^3}{12\pi^2M_\mathrm{pl}^6V_{,\phi}^2}$. The spectral index $n_s$ and the tensor-scalar-ratio $r$ are $n_s\simeq1-\left(6\epsilon_V-2\eta_V\right)$ and $r\simeq{16\epsilon_V} $, respectively, where $\epsilon_V=\frac{1}{2}M_\mathrm{pl}^2 \big(\frac{V_{,\phi}}{V} \big)^2\simeq \epsilon$,  and $\eta_V={M_\mathrm{pl}^2} \frac{V_{,\phi\phi}}{V}\simeq \epsilon+\delta_\phi$.
To be consistent with the CMB observations, we choose the potential of the inflaton field to be the Starobinsky potential~\cite{Starobinsky1980}
\begin{align}\label{Potential}
 {V}(\phi)=\Lambda^{4}\left[1-\exp \left(-\sqrt{2 / 3} \phi / M_{\mathrm{pl}}\right)\right]^{2},
\end{align}
where $\Lambda$ is a constant. 

\section{parametric resonance}
\label{sec3}
Since the parametric resonance occurs deep inside the Hubble horizon $\left(c_s k \gg aH\right)$, the Eq.~(\ref{sasaki2}) can be simplified to be 
\begin{eqnarray}\label{sasaki3}
\ddot{u}_{k}+c_{s}^{2}\frac{k^{2}}{a^{2}}u_{k}\approx 0. 
\end{eqnarray}
 Considering the slow-roll conditions given in Eq.~(\ref{SLC}), $w\ll \frac{M^2_\mathrm{pl}}{3H^2}$ and $\phi_c\ll \phi$ during inflation, we  find that the background equations (Eqs.~(\ref{br2}, \ref{br3})) can be reduced to
\begin{align}\label{hbr1}
 \dot{H}\approx \frac{1}{2M_{\mathrm{pl}}^{4}}H\dot{\phi}^{3}{\theta}_{,\phi}(\phi),~~~~\ddot{\phi}\approx - \frac{3}{2M_{\mathrm{pl}}^{2}}H^{2}\dot{\phi}^{2}{\theta}_{,\phi}(\phi),
\end{align}
  and then the sound speed square of the curvature perturbations can be simplified to be   
 \begin{align}\label{cs2}
c_{s}^{2}\approx  1+\delta c_s=1+\frac{3w^{2}H\dot{\phi }^{3}}{2\phi _{c}M_{\mathrm{pl}}^{6}} \sin{\frac{2\phi }{\phi _{c} } }.
\end{align}
 We find that $\delta c_s$  oscillates around zero and satisfies $|\delta c_s|\ll1$. So, the sound speed could exceed the speed of light.  However,  it has been found that the superluminal sound speed will not result in the causal paradoxes when the scalar field is non-trivial~\cite{Babichev2008, Armendariz2001, Mukhanov2006, Armendariz1999,Babichev2006, Kang2007, Armendariz2005}. This suggests that  there may  be no violation of causality  from the superluminal sound speed for the model considered in this  paper.  If a different coupling function, i.e. $\theta(\phi)\sim \sin^2(\phi)$,  is chosen, we find that the subliminal oscillation of the sound speed square is possible. 
 Substituting Eq.~(\ref{cs2}) into Eq.~(\ref{sasaki3}), one can obtain 
\bea \label{madiu3}
\ddot{u}_{k}+\left(\frac{k^{2}}{a^{2}}+\frac{3w^{2}H_{s}\dot{\phi}_{s}^{3}k^{2}}{2\phi_{c}M_{\mathrm{pl}}^{6}a^{2}}\sin\frac{2\phi }{\phi _{c}} \right) u_{k}\approx 0.
\eea

We assume that the inflaton field evolves from $\phi_s$ to $\phi_e$ during a short time, which indicates that during this short time the evolution of $\phi$ can be expressed approximately  as $\phi \approx  \phi_{s}+\dot{\phi}_{s}\left(t-t_{s}\right)$, where $t_s$ is the time when $\phi=\phi_s$. Setting $ x=\frac{\phi }{\phi_{c} }  -\frac{\pi }{4} =\frac{\dot{\phi}_{s}}{\phi _{c}}t +\frac{(\phi_{s}-\dot{\phi }_{s}t_{s} )}{\phi_{c} }-\frac{\pi }{4}$ and $k_{c}=\frac{\left | \dot{\phi}_{s}  \right | }{\phi_{c} } $, we find that  the Eq.~(\ref{madiu3}) can be transformed  into the form of the Mathieu equation
\begin{align}\label{madiu0}
\frac{d^{2} u_{k}}{d x^{2}}+\left[A_{k}(x)-2 q \cos 2 x\right] u_{k}=0,
\end{align}
where
\begin{align}\label{madiu1}
A_{k}(x) \equiv \frac{k^{2}}{k_{c}^{2} a^{2}}, ~~~q\equiv -\frac{3w^{2}H_{s}\dot{{\phi }_{s}}\phi _{c}k^{2} }{4M_{\mathrm{pl}}^{6}a^{2}}=2 C\frac{k^2}{k_c^2 a^2}.
\end{align}
 For the Mathieu equation, the resonant bands are close to narrow regions near $A_{k}(x)\simeq n^{2} ~\left(n=1,2,3... \right)$. The width of   each resonant band is $\Delta k\sim q^{n}$. If $0< q\ll 1$, the resonance in the first resonant band $(n=1)$ is the most violent. Therefore, we only consider the influence of the first resonant band on $u_{k}$. % When the $k$-modes satisfy  $1-q<A_{k}(x)<1+q$,  the perturbations of these modes can enter the first resonance band and the parametric resonance occurs. 
%Since   $A_{k}(x)$ varies with $x$, the time that the $k$ mode  stays in the resonant band is finite, which is given by
%\begin{equation}
%\label{eq:T_k}
%\begin{aligned}
%T(k) = \begin{cases}
%0, & k \leq k_s \sqrt{1-q}, \\
%\min(t_e, t_{ek}) - \max(t_s, t_{sk}), & k_s \sqrt{1-q} < k < k_e \sqrt{1+q}, \\
%0, & k \geq k_e \sqrt{1+q}.
%\end{cases}
%\end{aligned}
%\end{equation}
%Here $k_{s}=k_{c}a_{s}$ and $k_{e}=k_{c}a_{e}$ with $a_s$ and $a_e$ being the value of the scale factor when the inflaton field equals $\phi_s$ and $\phi_e$,  respectively. In Eq.~(\ref{eq:T_k}), $t_e$ and $t_s$ satisfy $a(t_e)=a_e$ and $a(t_s)=a_s$, respectively, and  $t_{sk}$ and $t_{ek}$ represent,  respectively,  the time of $k$ mode entering and leaving the resonant band, which satisfy $A_{k}(t_{sk})=1+q$ and $A_{k}(t_{ek})=1-q$. Thus, the resonant amplified width of the power spectrum is $\Delta k=k_e-k_s$, which is determined by $k_c$, $\phi_e$ and $\phi_s$. 
 
For the first instability, the  Floquet index $\mu_k$ which describes the rate of exponential growth has the form
\begin{align}\label{mu} 
 \mu_{k}(t)=\Re\left(\sqrt{\left(\frac{q}{2}\right)^{2}-\left(\frac{k}{k_{c} a}-1\right)^{2}}\right)=\Re\left(\sqrt{\left(C\frac{k^2}{k_c^2 a^2}\right)^{2}-\left(\frac{k}{k_{c} a}-1\right)^{2}}\right).
 \end{align}
Here $\Re$ refers to taking the real part.  Then, we find that the resonance occurs in a narrow band 
 \begin{eqnarray}\label{k}
k_- < \frac{k}{a}<  k_+,
\end{eqnarray}
where $k_-=k_c(1-|C|)$ and $k_+=k_c(1+|C|)$. In obtaining the expressions of  $k_\pm$,  we have used the condition $|C|\ll1$ derived from $0<q\ll1$. Since Eq.~(\ref{k}) is time dependent,  the duration that the $k$ mode  stays in the resonant band is finite, which is given by $T_{in} (k)=\min(t_e, t_{F}) - \max(t_s, t_{I})$ with $k$ satisfying $k_- a_s < k < k_+ a_e$, 
where subscripts $s$ and $e$ represent the moment that the inflaton field equals $\phi_s$ and $\phi_e$,  respectively,   and  $t_{I}$ and $t_{F}$ represent,  respectively,  the time of $k$ mode entering and leaving the resonant band. Thus, the resonant amplified width of the power spectrum is $\Delta k=k_+a_e-k_-a_s$, which is determined mainly by $k_c$, $\phi_e$ and $\phi_s$.

   During the parametric resonance, the curvature perturbations will be enhanced exponentially 
 \begin{align}\label{muk} 
\mathcal{A}(k) \equiv\left|\frac{u_{k}\left(t_{F}\right)}{u_{k}\left(t_{I}\right)} \right| \simeq \exp \left(\int_{t_{I}}^{t_{F}} \mu_{k}(t) k_{c} d t\right).
  \end{align}
  Defining $\mathcal{B}_{k}(t)=\frac{k }{k_{c} a}$, the above integral can be re-expressed as 
\begin{eqnarray}
\mathcal{A}_{k}\left(\mathcal{B}_{k}\left(t_I \right), \mathcal{B}_{k}\left(t_F\right)\right) \simeq \exp \left(-\frac{k_{c}}{H_{s}} \int_{\mathcal{B}_{k}\left(t_I\right)}^{\mathcal{B}_{k}\left(t_F\right)} \sqrt{\left(C \mathcal{B}_{k}^{2}\right)^{2}-\left(\mathcal{B}_{k}-1\right)^{2}} \frac{d \mathcal{B}_{k}}{\mathcal{B}_{k}}\right).
\end{eqnarray}

The amplified modes can be divided into three groups: (1) the modes entering the band before $t_s$; (2) the modes entering the band after $t_s$ and exiting before $t_e$; (3) the modes exiting the band after $t_e$. For these three groups, the  wavenumber satisfies:  $k_- a_s <k\leq k_+ a_s$, $k_+ a_s <k< k_- a_e$, and $k_- a_e  \leq k < k_+ a_e $,   respectively. The corresponding $\mathcal{B}_{k}(t_I)$ and $\mathcal{B}_{k}(t_F)$ can be calculated as 
\begin{eqnarray}
&&\mathcal{B}_{k}(t_I)=\frac{k}{k_ca_s}, \quad \mathcal{B}_{k}(t_F)=\frac{k_-}{k_c} \quad \mathrm{for}\quad k_- a_s <k\leq k_+ a_s, \\
&&\mathcal{B}_{k}(t_I)=\frac{k_+}{k_c}, \quad \mathcal{B}_{k}(t_F)=\frac{k_-}{k_c} \quad \mathrm{for}\quad  k_+ a_s <k< k_- a_e, \\
&&\mathcal{B}_{k}(t_I)=\frac{k_+}{k_c}, \quad \mathcal{B}_{k}(t_F)=\frac{k}{k_c a_e} \quad \mathrm{for}\quad  k_- a_e  \leq k< k_+ a_e. 
\end{eqnarray}
Apparently, for the second group: $k_+ a_s <k< k_- a_e$,  $\mathcal{B}_{k}(t_I)$ and $\mathcal{B}_{k}(t_F)$ are independent of $k$, which results in that  $\mathcal{A}_{k}$ is independent of $k$ for  this group.

\begin{figure*}
	\centering		\includegraphics[width=0.45\linewidth]{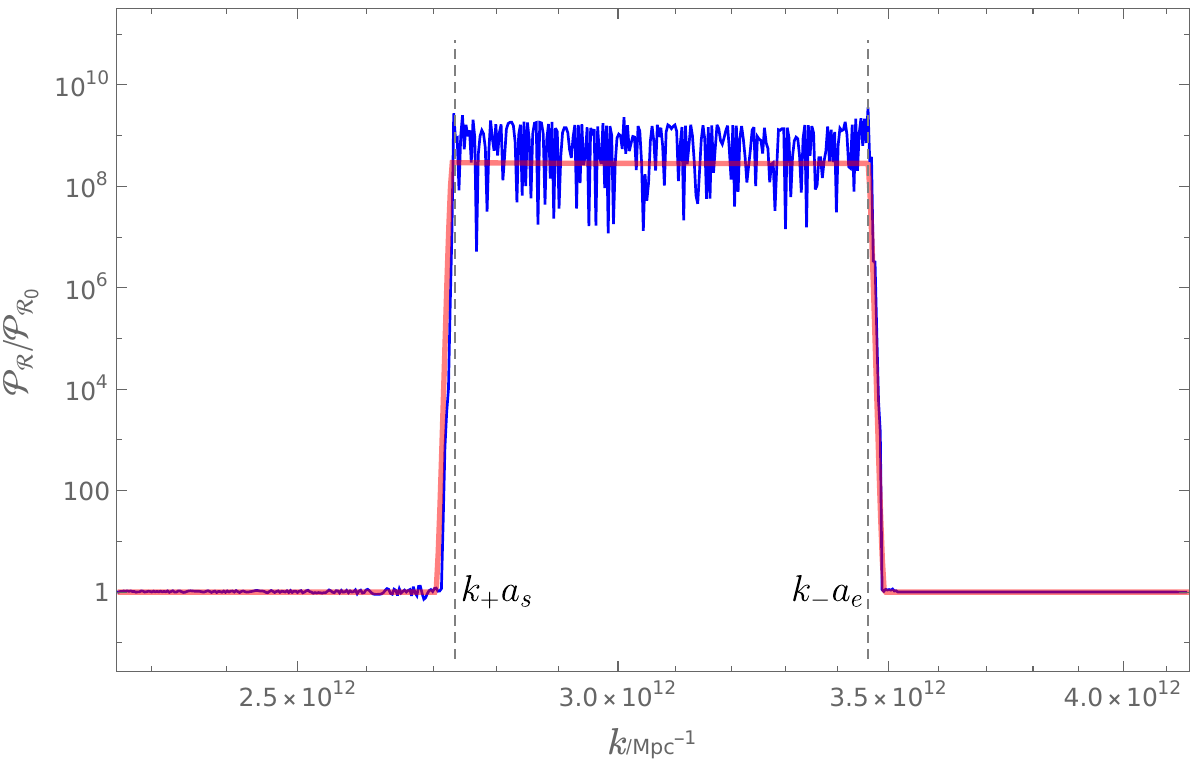}
		\caption{\label{AK}
		The evolution of $\mathcal{P}_\mathcal{R}/\mathcal{P}_{\mathcal{R}_0}$ with $k$. The blue and red lines represent the numerical results  from Eq.~(\ref{sasaki2}) and the approximate ones given in Eq.~(\ref{ARs}).  The parameters are set to be  $w=2.36\times10^{8}$, $\Lambda=3.14 \times10^{-3} M_\mathrm{pl}$,  $\phi_{s}=4.987 M_\mathrm{pl} $, $\phi_{e}=4.98 M_\mathrm{pl}$ and $\phi_{c}=9.98\times10^{-8} M_\mathrm{pl}$. } 
\end{figure*}

The enhanced  power spectrum of the curvature perturbations can be expressed approximately  as 
\begin{eqnarray}\label{ARs}
\mathcal{P}_\mathcal{R}\left(k\right) \approx\mathcal{A}_k^{2}\mathcal{P}_{\mathcal{R}_{0}}\left(k\right).
\end{eqnarray}
In the region of $k_+ a_s <k< k_- a_e$,  since $\mathcal{A}_k$ is independent of $k$, the enhanced power spectrum in this region  will have a plateau, which can be seen in Fig.~(\ref{AK}),  where we plot the  evolution of $\mathcal{P}_\mathcal{R}/\mathcal{P}_{\mathcal{R}_0}$ with $k$. In this figure the blue and red lines represent the numerical results  from Eq.~(\ref{sasaki2}) and the approximate  ones  given in Eq.~(\ref{ARs}), respectively. It is easy to see that the approximate results are consistent well with the numerical ones, and the power spectrum can be enhanced by several orders.   These enhanced curvature perturbations can lead to the generation of significant gravitational quadrupole moments during inflation, which  will emit GWs. However, this issue is beyond the scope of the present paper and is left to be investigated  in the future.

\begin{figure*}
	\centering		\includegraphics[width=0.45\linewidth]{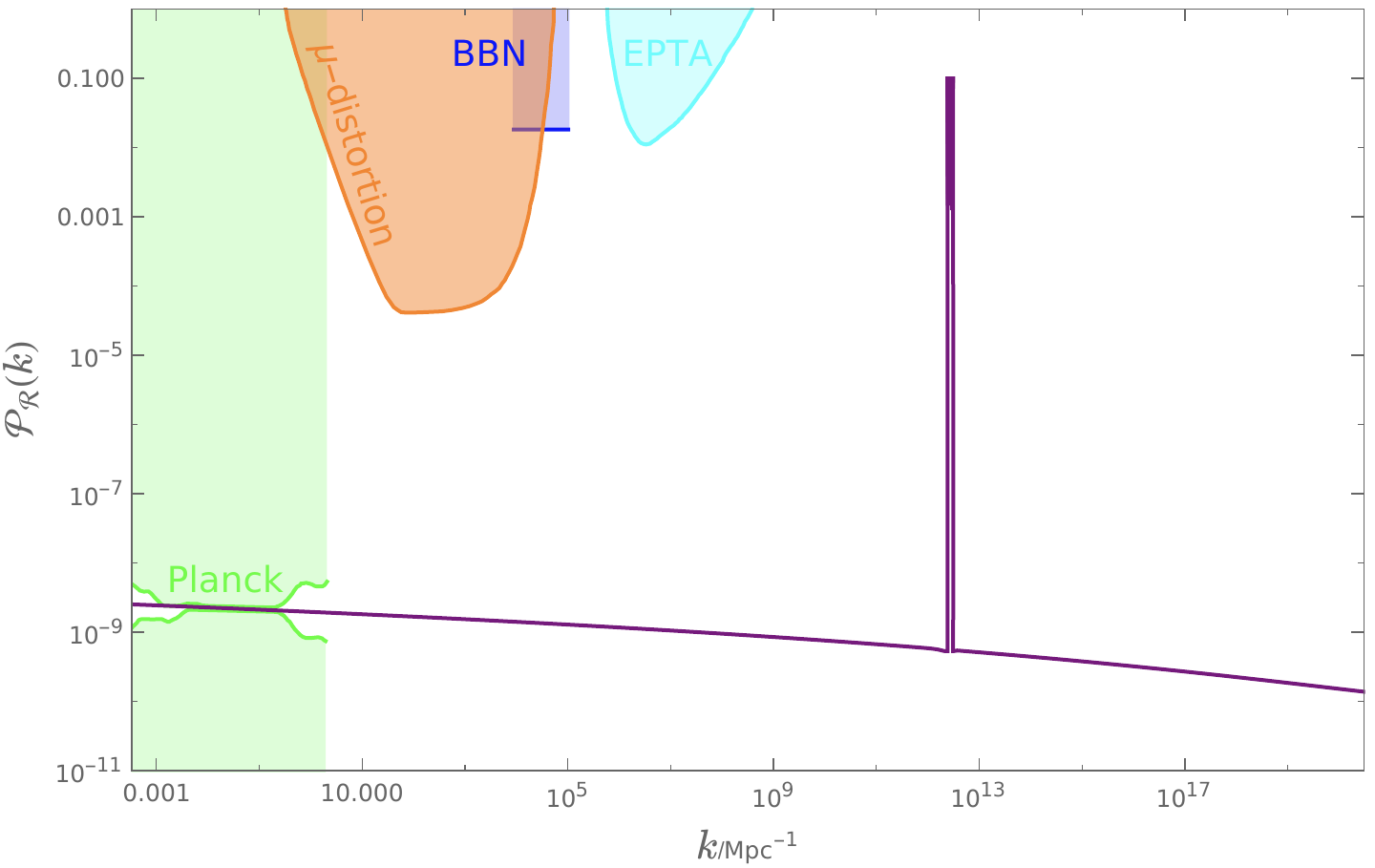}
		\caption{\label{PR}The power spectrum  of the curvature perturbations. The green shaded region is excluded from the current CMB observation \cite{Aghanim2020}. The orange- and blue-shaded regions are excluded by the $\mu$ distortion of CMB \cite{D.J.Fixsen} and the effect on the $n-p$ ratio during big-bang nucleosynthesis (BBN) \cite{K.Inomata2016}, respectively. Cyan shaded region indicate the limitations of current PTA observations in the power spectrum \cite{K.Inomata2019}.}
\end{figure*}

Figure (\ref{PR}) shows the  power spectrum of the curvature perturbations  from numerical calculation, which indicates clearly that the power spectrum is compliant with the CMB observations at the CMB scale, and it can be amplified to generate a sizable amount of PBHs at scales smaller than the CMB one.

\section{PBHs}
\label{sec4}
When the sufficiently large curvature perturbations re-enter the Hubble horizon during the radiation-dominated period, the gravity of the high-density regions will overcome the radiation pressure and lead to the formation of  PBHs. The  PBH mass has the following relationship with  $k$: 
\begin{align}\label{23}
 M(k)=\gamma\frac{4\pi M_\mathrm{pl}^2}{H} \simeq M_{\odot}\left(\frac{\gamma}{0.2}\right)\left(\frac{g_{*}}{10.75}\right)^{-\frac{1}{6}}\left(\frac{k}{1.9 \times 10^{6}~ \mathrm{Mpc}^{-1}}\right)^{-2} \;.
\end{align} 
Here $\gamma $ is the ratio of the mass of PBH to the total mass of the Hubble horizon when the PBH is formed.  It represents the effective collapse rate, and its specific value is related to the details of gravitational collapse. In our analysis we set  $ \gamma \simeq(1 / \sqrt{3})^{3}$ \cite{Carr1975}.  In Eq.~(\ref{23}), $M_\odot$ represents the solar mass, and  $g_{*}$  is the number of degrees of freedom of the relativistic particle at the time of the PBH formation. Assuming that the PBHs form in the radiation-dominated period, we can set $g_{*} = 106.75$. 

\begin{figure*}
	\centering
	\includegraphics[width=0.45\linewidth]{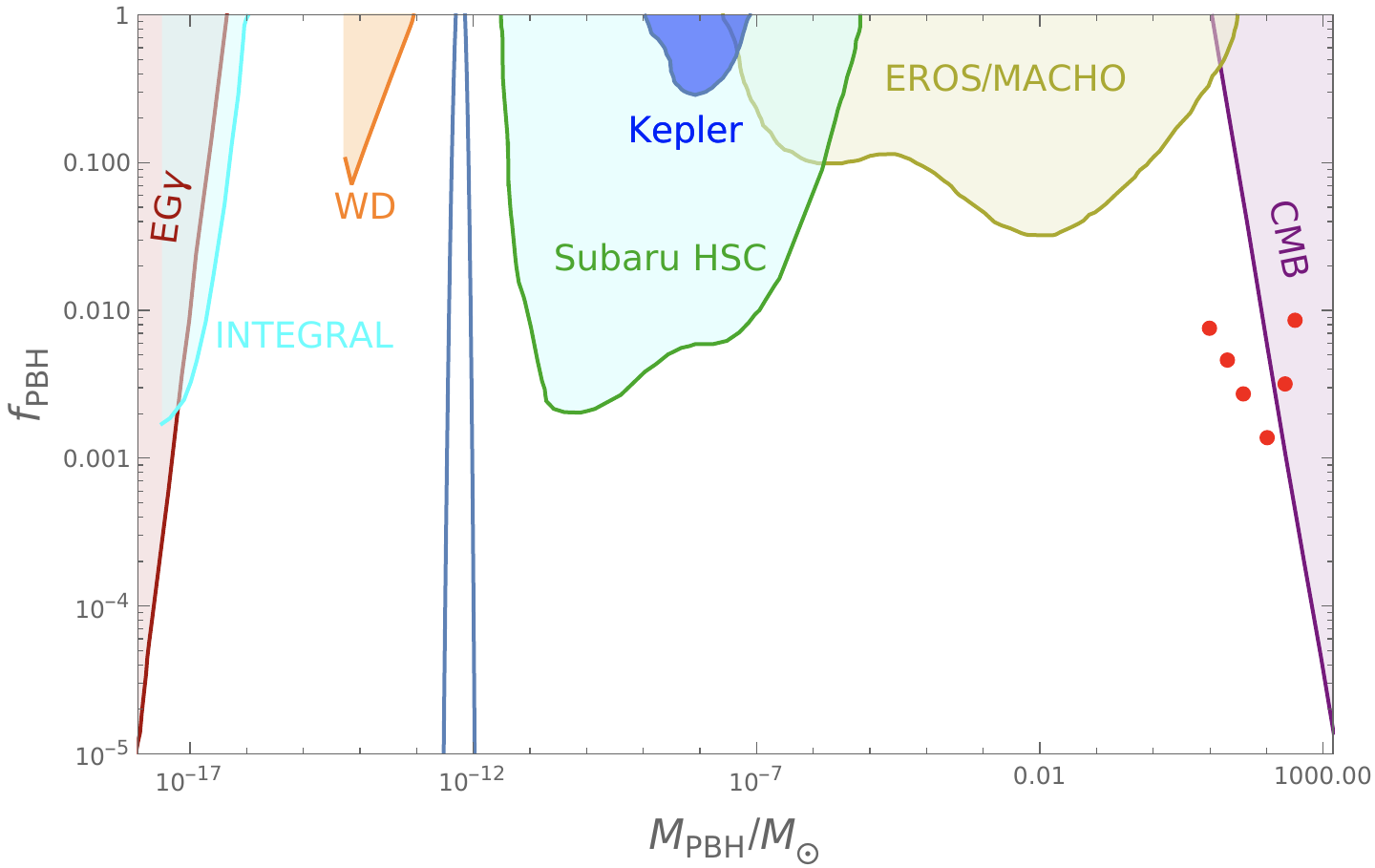}
	\caption{\label{PBH} The mass spectrum of PBHs. The colored regions are ruled out by observations including the extragalactic gamma ray background ($\mathrm{EG}\gamma$~\cite{B.J.Carr2010} ), the galactic center 511 keV gamma-ray line (INTEGRAL)~\cite{R.Laha}, the white dwarfs explosion (WD)~\cite{P.W.Graham2015}, the microlensing events with Subaru HSC (Subaru HSC)~\cite{H.Niikura2019}, with the Kepler satellite (Kepler)~\cite{K.Griest}, with EROS/MACHO (EROS/MACHO)~\cite{P.Tisserand}, and the  CMB~\cite{V. Poulin}. }
\end{figure*}

  Based on the Press-Schechter theory \cite{Tada2019,Young2014}, the production rate of  PBHs with mass $M(k)$ is
\bea
\beta(M)=\int_{\delta_{c}} \frac{d \delta}{\sqrt{2 \pi \sigma^{2}(M)}} e^{-\frac{\delta^{2}}{2 \sigma^{2}(M)}}=\frac{1}{2} \operatorname{erfc}\left(\frac{\delta_{c}}{\sqrt{2 \sigma^{2}(M)}}\right)
\eea
after assuming  that the probability distribution function of the disturbance obeys the Gaussian distribution. Here $\mathrm{erfc}$ is the complementary error function, and $\delta_c$  is the threshold for the relative density perturbation of the PBH formation, which is chosen to be $\delta _{c} \simeq 0.4$ \cite{Musco2013,Harada2013} in our calculation of the PBH abundance. The variance $\sigma^{2}(M)$ represents the coarse-grained density contrast with the smoothing scale $k$, and it takes the form
 \bea \sigma^{2}(M(k))=  \frac{16}{81} \int d \ln q \; W^{2}\left(q k^{-1}\right) \left(q k^{-1}\right)^{4} \mathcal{P}_{\mathcal{R}}(q) \; .
\eea
   Here $W$ is  the window function. We find that the PBH mass spectrum can be  obtained from the following equation 
\begin{align}
f(M)  \equiv \frac{1}{\Omega_{\mathrm{DM}}} \frac{d \Omega_{\mathrm{PBH}}}{d \ln M} \simeq \frac{\beta(M)}{1.84 \times 10^{-8}}\left(\frac{\gamma}{0.2}\right)^{\frac{3}{2}}\left(\frac{10.75}{g_{*}}\right)^{\frac{1}{4}}\left(\frac{0.12}{\Omega_{\mathrm{DM}} h^{2}}\right)\left(\frac{M}{M_{\odot}}\right)^{-\frac{1}{2}}.
\end{align}
Here $\Omega_{\mathrm{DM}}$ represents the current dark matter density parameter and $h$ is the reduced Hubble constant.  While $\Omega_{\mathrm{DM}}h^{2}$ is constrained to be $\Omega_{\mathrm{DM}}h^{2} \simeq  0.12$ by the Planck 2018 observations~\cite{Aghanim2020}. 
 We show the numerical results of the  PBH mass spectrum in Fig.~(\ref{PBH}) and find that the PBHs can make up most of  dark matter since $\frac{\Omega_\mathrm{PBH}}{\Omega_\mathrm{DM}}\simeq 0.99$. 
%\textbf{We present the numerical results of the PBHs mass spectrum in Fig.~(\ref{PBH}), where it can be observed that $f_{\mathrm{PBH}} > 1$. This setting is chosen to allow PBHs to constitute a significant portion of dark matter since the integration yields $\frac{\Omega_\mathrm{PBH}}{\Omega_\mathrm{DM}}\simeq 0.99$.}

\section{SIGWs}
\label{sec5}
 
\begin{figure*}
	\centering
	\includegraphics[width=0.45\linewidth]{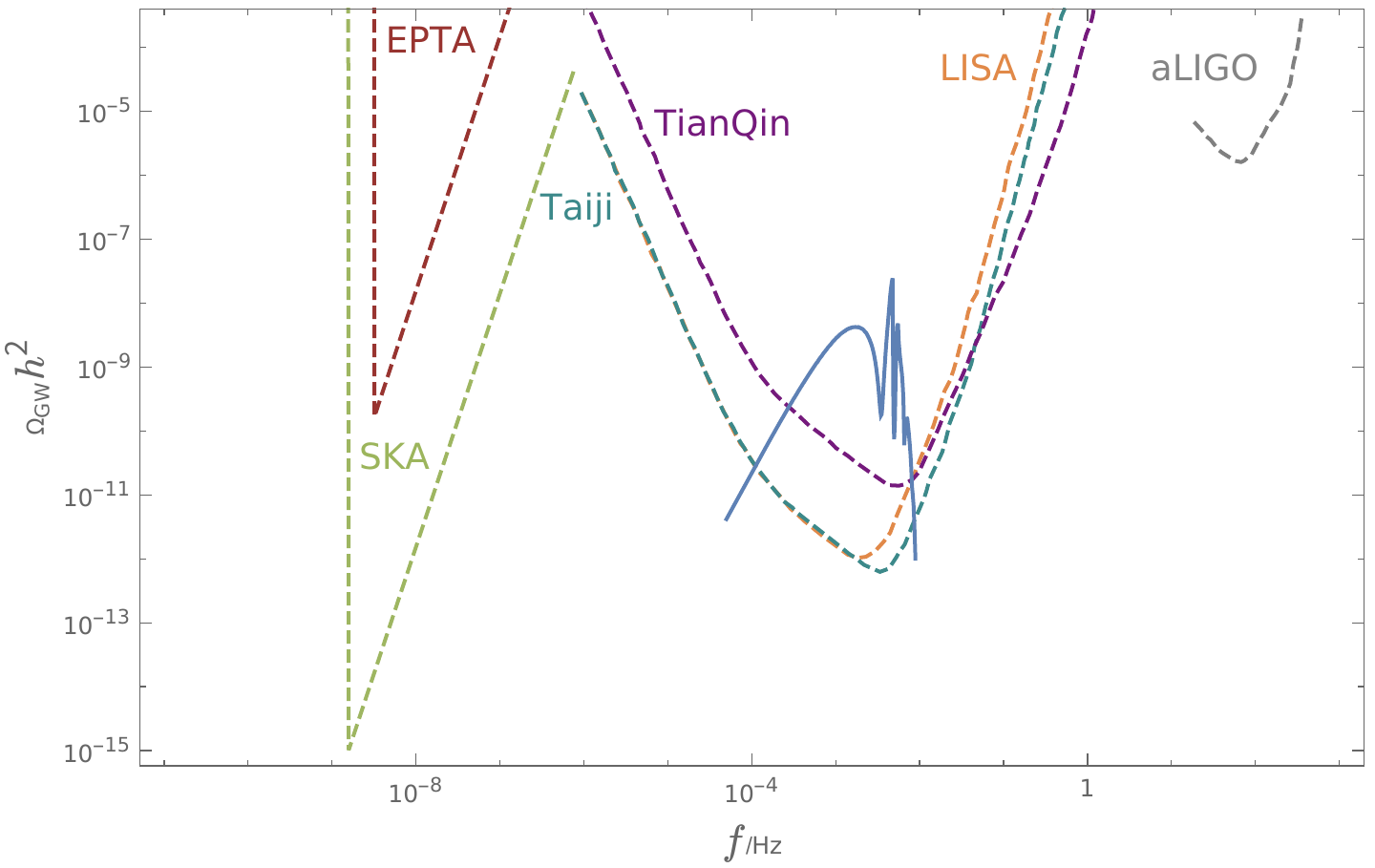}
		\caption{\label{GW} The current energy spectrum (solid blue line) of SIGWs. The various dotted lines represent the sensitive intervals of different gravitational wave detectors including  SKA \cite{ska}, EPTA \cite{epta}, TAIJI \cite{taiji}, TIANQIN \cite{tianqin}, LISA \cite{lisa}, and ALIGO \cite{aligo}.  }
\end{figure*}

Associated with the formation of PBHs, which are assumed to be generated in the radiation-dominated era, the large metric scalar perturbations become an important GW source and radiate the  observable SIGWs.  The second-order tensor perturbations $h_{ij}$ satisfy the equation: 
\begin{eqnarray}\label{hij}
h_{ij}''+2\mathcal{H}h_{ij}'-\nabla^2 h_{ij}=-4\mathcal{T}_{ij}^{lm}S_{lm},
\end{eqnarray}
 where a prime denotes the derivative with respect to the conformal time, $\mathcal{H}\equiv a'/a$,  $\mathcal{T}_{ij}^{lm}$ is the transverse-traceless projection operator, and   \begin{align}
 S_{i j}=4 \Psi \partial_{i} \partial_{j} \Psi+2 \partial_{i} \Psi \partial_{j} \Psi-\frac{1}{\mathcal{H}^{2}} \partial_{i}\left(\mathcal{H} \Psi+\Psi^{\prime}\right) \partial_{j}\left(\mathcal{H} \Psi+\Psi^{\prime}\right)
 \end{align}
 is the GW source term~\cite{Ananda2007,Baumann2007}. 
 Here $\Psi$ is the  metric scalar perturbation. In the radiation-dominated era, the  evolution of $\Psi$ is $\Psi_{k}(\eta)=\psi_{k} \frac{9}{(k \eta)^{2}}\left(\frac{\sin (k \eta / \sqrt{3})}{k \eta / \sqrt{3}}-\cos (k \eta / \sqrt{3})\right)$~\cite{Baumann2007}, 
 where $\psi_{k}$ is the primordial perturbation, which relates with the power spectrum of the primordial curvature perturbations through 
\begin{align}
\left\langle\psi_{\mathbf{k}} \psi_{\tilde{\mathbf{k}}}\right\rangle=\frac{2 \pi^{2}}{k^{3}}\left(\frac{4}{9} \mathcal{P}_{\mathcal{R}}(k)\right) \delta(\mathbf{k}+\tilde{\mathbf{k}}).
\end{align}
 Solving Eq.~(\ref{hij}), one can obtain the GW energy density for each logarithmic interval $k$~\cite{Kohri2018}:
 \begin{align}
\Omega_{\mathrm{GW}}\left(\eta_{c}, k\right)=& \frac{1}{12} \int_{0}^{\infty} d v \int_{|1-v|}^{|1+v|} d u\left(\frac{4 v^{2}-\left(1+v^{2}-u^{2}\right)^{2}}{4 u v}\right)^{2} \mathcal{P}_{\mathcal{R}}(k u) \mathcal{P}_{\mathcal{R}}(k v) \nonumber\\
&\times \left(\frac{3}{4 u^{3} v^{3}}\right)^{2}\left(u^{2}+v^{2}-3\right)^{2} \times \bigg \{\left[-4 u v+\left(u^{2}+v^{2}-3\right) \ln \left|\frac{3-(u+v)^{2}}{3-(u-v)^{2}}\right|\right]^{2} \nonumber
\\ &+\pi^{2}\left(u^{2}+v^{2}-3\right)^{2} \Theta(v+u-\sqrt{3})\bigg\}\, ,
\end{align}
where %$u=p/k$ and $v=|\boldsymbol{k}-\boldsymbol{p}|/k$,  and the 
$\eta_{c}$ represents the time when $\Omega_{\mathrm{GW}}$ stops to grow.
The current energy density spectrum of SIGWs can be expressed as \cite{Kohri2018, K.Inomata2019}
\begin{align}
\Omega_{\mathrm{GW}, 0} h^{2}=0.83\left(\frac{g_{*}}{10.75}\right)^{-1 / 3} \Omega_{\mathrm{r}, 0} h^{2} \Omega_{\mathrm{GW}}\left(\eta_{c}, k\right),
\end{align}
where $\Omega_{\mathrm{r}, 0} h^{2}$ is the current density parameter of radiation which is set to be $4.2\times 10^{-5}$.
The current frequency $f$ of  SIGWs is related to  $k$ through
\begin{align}
f=1.546 \times 10^{-15} \frac{k}{1 \mathrm{Mpc}^{-1}} \mathrm{~Hz}.
\end{align}
In Fig.~(\ref{GW}),  we show the current  energy spectrum of SIGWs. One can see that the SIGWs possess a multi-peak structure and  may be detected by the future GW projects including LISA, Taiji and TianQin.

 \section{conclusions}
 \label{conclusion}
 
In order to produce a sizable amount of PBHs, the amplitude of the power spectrum of the small-scale  curvature perturbations must be enhanced  by about 7 orders of magnitude compared to that at  the CMB scale. In inflation models in which the inflaton field couples derivatively with  gravity, it has been found that the curvature perturbations can be amplified through the gravitationally enhanced friction  mechanism~\cite{Fuchengjie2019}.  In this paper, we find  that, if there is a weak,  transient and periodical derivative  coupling between  the inflaton field and  gravity,  the  sound speed square of the curvature perturbations becomes a periodic function, which results in that the equation of the curvature perturbations  in the sub-horizon limit can be transformed into the form of   the Mathieu equation.  Thus, for some $k$-modes, the parametric resonance will amplify their fluctuations. These amplified fluctuations are stretched to be super-horizon by the inflation and then   the  power spectrum will be enhanced  at scales smaller than the CMB one.  When the enhanced curvature perturbations re-enter the horizon during radiation-dominated era, they will lead to the formation of PBHs, which can explain most of  dark matter.  Associated  with the generation  of PBHs, the large scalar perturbations will  radiate the  observable SIGWs. We demonstrate that the current energy spectrum of the SIGWs has a multi-peak structure, which is different from that in the  inflation model with the gravitationally enhanced friction~\cite{fuchengjie2020}, and it can be  detected by future GW projects including LISA, Taiji and TianQin. Therefore,  the future detection of  SIGWs will help us to distinguish  different mechanisms of enhancing curvature perturbations at small scales. Finally, 
it is worth noting that when loop corrections are considered the perturbation theory will be broken in the cases of the amplification of the primordial curvature perturbations due to the decrease of the inflaton's rolling speed~\cite{Kristiano2022} and   the parametric resonance from the oscillating potential~\cite{Inomata2023}. Whether the loop corrections will break the perturbation theory in the scenario considered in the present paper needs to be studied separately since the gravity, which couples derivatively with the inflaton field, is different from the theory of general relativity.

\begin{acknowledgments}
We appreciate very much the insightful comments and helpful suggestions by anonymous referees. This work is supported by the National Key Research and Development Program of China Grant No.~2020YFC2201502, and  by the National Natural Science Foundation of China under Grants No.~12275080 and No.~12075084. 
\end{acknowledgments}


\begin{thebibliography}{99}

 
 \bibitem{Zeldovich}
B. Ya Zel\textquotesingle dovich and I. D. Novikov,
\href{https://ui.adsabs.harvard.edu/abs/1966AZh....43..758Z/abstract}
{Soviet Astron.AJ (Engl. Transl.), 	\textbf{9}, 602 (1967)}.
 
\bibitem{Hawking}
S. Hawking, 
\href{https://doi.org/10.1093/mnras/152.1.75}
{Mon. Not. Roy. Astron. Soc. \textbf{152}, 75 (1971)}.

\bibitem{Carr}
B. J. Carr and S. W. Hawking, 
\href{https://doi.org/10.1093/mnras/168.2.399}
{Mon. Not. Roy. Astron. Soc. \textbf{168}, 399 (1974)}.

\bibitem{Meszaros}
P. Meszaros, 
\href{http://pascal-francis.inist.fr/vibad/index.php action=getRecordDetail&idt=PASCAL7630047603}
{Astron. Astrophys. \textbf{37}, 225 (1974)}.

\bibitem{Carr1975} 
B. J. Carr, 
\href{http://dx.doi.org/10.1086/153853}
{Astrophys. J. \textbf{201}, 1 (1975)}.

\bibitem{Khlopov} 
M. Y. Khlopov, B. A. Malomed, and Y. B. Zeldovich, 
\href{https://doi.org/10.1093/mnras/215.4.575}
{Mon. Not. Roy. Astron. Soc. \textbf{215}, 575 (1985)}.


\bibitem{Ozsoy2023} 
O. \"Ozsoy and G. Tasinato, 
\href{https://doi.org/10.3390/universe9050203}
{Universe \textbf{2023}, 9(5), 203}.

\bibitem{Choudhury2023f}S. Choudhury, K. Dey, and A. Karde, \href{https://doi.org/10.48550/arXiv.2311.15065}{arXiv: 2311.15065}.
 
 
\bibitem{lg1}
S. Bird, I. Cholis, J. B. Mu\~{n}oz, Y. Ali-Ha\"{i}moud, M. Kamionkowski, E. D. Kovetz, A. Raccanelli, and A. G. Riess,
\href{https://doi.org/10.1103/PhysRevLett.116.201301}
{Phys. Rev. Lett. \textbf{116}, 201301 (2016)}.


\bibitem{lg2}
S. Clesse and J. Garc\'{i}a-Bellido,
\href{https://doi.org/10.1016/j.dark.2016.10.002}	
{Phys. Dark Univ. \textbf{15}, 142 (2017)}.


\bibitem{lg3}
M. Sasaki, T. Suyama, T. Tanaka, and S. Yokoyama,
\href{https://doi.org/10.1103/PhysRevLett.121.059901}
{Phys. Rev. Lett. \textbf{117}, 061101 (2016)}.


\bibitem{lg4}
B. Carr, F. Kuhnel, and M. Sandstad,
\href{https://doi.org/10.1103/PhysRevD.94.083504}
{Phys. Rev. D \textbf{94}, 083504 (2016)}.


 \bibitem{P.Mroz2017}
P. Mr\'oz, A. Udalski,  J. Skowron, et al.,
\href{https://doi.org/10.1038/nature23276}
{Nature (London) \textbf{548}, 183 (2017)}.


\bibitem{H.Niikura2019}
H. Niikura, M. Takada, S. Yokoyama, T. Sumi, and S. Masaki,
\href{https://doi.org/10.1103/PhysRevD.99.083503}
{Phys. Rev. D \textbf{99}, 083503 (2019)}.

  \bibitem{J.Scholtz2020}
J. Scholtz and J. Unwin,
\href{https://doi.org/10.1103/PhysRevLett.125.051103}
{Phys. Rev. Lett. \textbf{125}, 051103 (2020)}.

\bibitem{E.Witten}
E. Witten, 
\href{https://doi.org/10.48550/arXiv.2004.14192}
{arXiv: 2004.14192}.

\bibitem{A.Katz2018}
A. Katz, J. Kopp, S. Sibiryakov, and W. Xue,
\href{https://doi.org/10.1088/1475-7516/2018/12/005}
{J. Cosmol. Astropart. Phys. {12} (2018) 005}.


\bibitem{A.Barnacka2012}
A. Barnacka, J. F. Glicenstein, and R. Moderski,
\href{https://doi.org/10.1103/PhysRevD.86.043001}
{Phys. Rev. D \textbf{86}, 043001 (2012)}.

\bibitem{P.W.Graham2015}
P. W. Graham, S. Rajendran, and J. Varela,
\href{https://doi.org/10.1103/PhysRevD.92.063007}
{Phys. Rev. D \textbf{92}, 063007 (2015)}.

\bibitem{H.Niikura2019a}
H. Niikura, M. Takada, N. Yasuda, et al.,
\href{https://doi.org/10.1038/s41550-019-0723-1}
{Nat. Astron. \textbf{3}, 524 (2019)}.

\bibitem{lisa}
P. Amaro-Seoane,  H. Audley, S. Babak, et al. (LISA),
\href{https://doi.org/10.48550/arXiv.1702.00786}
{arXiv:1702.00786}.

\bibitem{taiji}
W. R. Hu and Y. L. Wu,
\href{https://doi.org/10.1093/nsr/nwx116}
{Natl. Sci. Rev. \textbf{4}, 685 (2017)}.

\bibitem{tianqin}
J. Luo,  L. S. Chen, H. Z. Duan, et al. (TianQin),
\href{https://doi.org/10.1088/0264-9381/33/3/035010}
{Class. Quant. Grav. \textbf{33}, 035010 (2016)}.	

\bibitem{pta1}
R. D. Ferdman,  R. van Haasteren, C. G. Bassa, et al.,
\href{https://doi.org/10.1088/0264-9381/27/8/084014}
{Class. Quant. Grav. \textbf{27}, 084014 (2010)}.

\bibitem{pta2}
G. Hobbs,  A.  Archibald2, Z. Arzoumanian, et al.,
\href{https://doi.org/10.1088/0264-9381/27/8/084013}
{Class. Quant. Grav. \textbf{27}, 084013 (2010)}.

\bibitem{pta3}
M.  A. McLaughlin,	
\href{https://doi.org/10.1088/0264-9381/30/22/224008}
{Class. Quant. Grav. \textbf{30}, 224008 (2013)}.

\bibitem{pta4}
G. Hobbs, %The Parkes Pulsar Timing Array,
\href{https://doi.org/10.1088/0264-9381/30/22/224007}
{Class. Quant. Grav. \textbf{30}, 224007 (2013)}.

 
% \bibitem{Y-H2023}
%Y-H. Yu, S. Wang,
%\href{https://doi.org/10.48550/arXiv.2303.03897}
%{arXiv: 2303.03897}.
% 
 
 
\bibitem{Aghanim2020}
N. Aghanim, Y. Akrami, M. Ashdown, et al. (Planck Collaboration),
\href{https://doi.org/10.1051/0004-6361/201833910}
{Astron. Astrophys. \textbf{641}, A6 (2020)}.

\bibitem{Yokoyama1998}
J. Yokoyama, 
\href{http://dx.doi.org/10.1103/PhysRevD.58.083510}
{Phys. Rev. D \textbf{58}, 083510 (1998).}

\bibitem{Choudhury2014}
S. Choudhury and A.  Mazumdar, 
\href{https://doi.org/10.1016/j.physletb.2014.04.050}
{Phys. Lett. B \textbf{733}, 270 (2014).}

\bibitem{Tada2019}
 Y. Tada and S. Yokoyama,
  \href{http://dx.doi.org/10.1103/PhysRevD.100.023537}
  {Phys. Rev. D \textbf{100}, 023537 (2019)}.

\bibitem{Mishra2020}
S. S. Mishra and V. Sahni,
 \href{http://dx.doi.org/10.1088/1475-7516/2020/04/007}
{J. Cosmol. Astropart. Phys.  {04} (2020) 007}.

\bibitem{Atal2020}
V. Atal, J. Cid, A. Escriva, and J. Garriga,
\href{https://doi.org/10.1088/1475-7516/2020/05/022}
{J. Cosmol. Astropart. Phys.  {05} (2020) 022}.

\bibitem{Ragavendra2021}
H. V. Ragavendra, P. Saha, L. Sriramkumar, and J. Silk,
\href{https://doi.org/10.1103/PhysRevD.103.083510}
{Phys. Rev. D \textbf{103}, 083510 (2021)}.

\bibitem{Bhaumik2020} 
N. Bhaumik and R. K. Jain, 
\href{http://dx.doi.org/10.1088/1475-7516/2020/01/037} 
{J. Cosmol. Astropart. Phys.  {01} (2020) 037}.

\bibitem{Motohashi2017}
H. Motohashi and W. Hu,
\href{https://doi.org/10.1103/PhysRevD.96.063503}
{Phys. Rev. D \textbf{96}, 063503 (2017)}.

\bibitem{Ezquiaga2018}
J. M. Ezquiaga, J. Garcia-Bellido, and E. Ruiz Morales,
\href{https://doi.org/10.1016/j.physletb.2017.11.039}
{Phys. Lett. B \textbf{776}, 345 (2018)}.

\bibitem{H. Di2018}
H. Di and Y. Gong,
\href{https://doi.org/10.1088/1475-7516/2018/07/007}
{J. Cosmol. Astropart. Phys. {07} (2018) 007}.

\bibitem{Ballesteros2018}
G. Ballesteros and M. Taoso,
\href{https://doi.org/10.1103/PhysRevD.97.023501}
{Phys. Rev. D \textbf{97}, 023501 (2018)}.

\bibitem{Dalianis2019}
I. Dalianis, A. Kehagias, and G. Tringas,
\href{https://doi.org/10.1088/1475-7516/2019/01/037}
{J. Cosmol. Astropart. Phys. {01} (2019) 037}.

\bibitem{Gao2018}
T. J. Gao and Z. K. Guo,
\href{https://doi.org/10.1103/PhysRevD.98.063526}
{Phys. Rev. D \textbf{98}, 063526 (2018)}.

\bibitem{Drees2021}
M. Drees and Y. Xu,
\href{https://doi.org/10.1140/epjc/s10052-021-08976-2}
{Eur. Phys. J. C \textbf{81}, 182 (2021)}.

\bibitem{C.Fu2020}
C. Fu, P. Wu, and H. Yu,
\href{https://doi.org/10.1103/PhysRevD.102.043527}
{Phys. Rev. D \textbf{102}, 043527 (2020)}.

\bibitem{Xu2020}
W. Xu, J. Liu, T. Gao, and Z. Guo,
\href{https://doi.org/10.1103/PhysRevD.101.023505}
{Phys. Rev. D \textbf{101},  023505 (2020)}.

\bibitem{Lin2020}
J. Lin, Q. Gao, Y. Gong, Y. Lu, C. Zhang, and F. Zhang, 
\href{https://doi.org/10.1103/PhysRevD.101.103515}
{Phys. Rev. D \textbf{101}, 103515 (2020)}.

\bibitem{Dalianis2021}
I. Dalianis and K. Kritos,
\href{https://doi.org/10.1103/PhysRevD.103.023505}
{Phys. Rev. D \textbf{103}, 023505 (2021)}.

\bibitem{Yi2021}
Z. Yi, Y. Gong, B. Wang, and Z. Zhu,
\href{https://doi.org/10.1103/PhysRevD.103.063535}
{Phys. Rev. D \textbf{103}, 063535 (2021)}.

\bibitem{Gao2021}
Q. Gao, Y. Gong, and Z. Yi, 
\href{https://doi.org/10.1016/j.nuclphysb.2021.115480}
{Nucl. Phys. B \textbf{969}, 115480  (2021)}.

\bibitem{Yi2021b}
Z. Yi, Q. Gao, Y. Gong, and Z. Zhu,
\href{https://doi.org/10.1103/PhysRevD.103.063534}
{Phys. Rev. D \textbf{103}, 063534 (2021)}.

\bibitem{TGao2021}
T. Gao and X. Yang,
\href{https://doi.org/10.1140/epjc/s10052-021-09269-4}
{Eur. Phys. J. C \textbf{81}, 494 (2021)}. 

\bibitem{Solbi2021}
M. Solbi and K. Karami,
\href{https://doi.org/10.1088/1475-7516/2021/08/056}
{J. Cosmol. Astropart. Phys. {08} (2021) 056}.

\bibitem{Gao2021b}
Q. Gao,
\href{https://doi.org/10.1007/s11433-021-1708-9}
{Sci. China Phys. Mech. Astron. \textbf{64}, 280411 (2021)}.

\bibitem{Solbi2021b}
M. Solbi and K. Karami, 
\href{https://doi.org/10.1140/epjc/s10052-021-09690-9}
{Eur. Phys. J. C \textbf{81}, 884 (2021)}.


\bibitem{Cai2021}
R. Cai, C. Chen,  and C. Fu,
\href{https://doi.org/10.1103/PhysRevD.104.083537}
{Phys. Rev. D \textbf{104},  083537 (2021)}.

\bibitem{Wang2021}
Q. Wang, Y. Liu, B. Su, and N. Li,
\href{https://doi.org/10.1103/PhysRevD.104.083546}
{Phys. Rev. D \textbf{104}, 083546 (2021)}.


\bibitem{Germani2017}
C. Germani and T. Prokopec,
\href{https://doi.org/10.1016/j.dark.2017.09.001}
{Phys. Dark Univ. \textbf{18}, 6 (2017)}.



\bibitem{Zheng2021}
R. Zheng, J. Shi, and T. Qiu,
\href{https://doi.org/10.1088/1674-1137/ac42bd}
{Chin. Phys. C \textbf{46}, 045103 (2022)}.

\bibitem{Teimoori2021a}
Z. Teimoori, K. Rezazadeh, M. A. Rasheed, and K. Karami,
\href{https://doi.org/10.1088/1475-7516/2021/10/018}
{J. Cosmol. Astropart. Phys. {10} (2021) 018}.


\bibitem{Fuchengjie2019}
C. Fu, P. Wu, and H. Yu,
\href{https://doi.org/10.1103/PhysRevD.100.063532}
{Phys. Rev. D \textbf{100}, 063532 (2019)}.

\bibitem{fuchengjie2020}
C. Fu, P. Wu, and H. Yu,
\href{https://doi.org/10.1103/PhysRevD.101.023529}
{Phys. Rev. D \textbf{101}, 023529 (2020)}.

\bibitem{Dalianis2020}
I. Dalianis, S. Karydas, and E. Papantonopoulos, 
\href{https://doi.org/10.1088/1475-7516/2020/06/040}
{J. Cosmol. Astropart. Phys. {06} (2020) 040}.

\bibitem{Teimoori2021}
Z. Teimoori, K. Rezazadeh, and K. Karami,
\href{https://doi.org/10.3847/1538-4357/ac01cf}
{Astrophys. J. \textbf{915}, 118 (2021)}.

\bibitem{Heydari2022}
S. Heydari, and K. Karami, 
\href{https://doi.org/10.1140/epjc/s10052-022-10036-2}
{ Eur. Phys. J. C \textbf{82}, 83 (2022)}.

\bibitem{Heydari2022b}
S. Heydari, and K. Karami, 
\href{https://doi.org/10.1088/1475-7516/2022/03/033}
{J. Cosmol. Astropart. Phys. {03} (2022) 033}. 

\bibitem{Karam2022}
A. Karam, N. Koivunen, E. Tomberg, V. Vaskonen, and H. Veermae, 
\href{https://doi.org/10.1088/1475-7516/2023/03/013}
{J. Cosmol. Astropart. Phys. {03} (2023) 013}.




 \bibitem{Bellido2017} 
J. Garc{\'i}a-Bellido and E. R. Morales, 
\href{https://doi.org/10.1016/j.dark.2017.09.007}
{Phys. Dark Univ. \textbf{18}, 47 (2017)}.





\bibitem{Pi2022} 
S. Pi and J. Wang, 
\href{https://doi.org/10.1088/1475-7516/2023/06/018}
{J. Cosmol. Astropart. Phys. {06} (2023) 018}.


\bibitem{Choudhury2023cc}
S. Choudhury, M. R. Gangopadhyay, and  M. Sami, 
\href{https://arxiv.org/pdf/2301.10000}
{arXiv: 2301.10000}.

\bibitem{Meng2022}
D. Meng, C. Yuan, and Q. Huang, 
\href{https://doi.org/10.1007/s11433-022-2095-5}
{Sci. China Phys. Mech. Astron. {\bf 66}, 280411 (2023)}.

\bibitem{Mu2022}
B. Mu, G. Cheng, J. Liu, and Z. Guo, 
\href{https://doi.org/10.1103/PhysRevD.107.043528}
{Phys. Rev. D \textbf{107}, 043528 (2023)}.

\bibitem{Kawaguchi2022}
R. Kawaguchi and S. Tsujikawa, 
\href{https://doi.org/10.1103/PhysRevD.107.063508}
{Phys. Rev. D \textbf{107}, 063508 (2023)}.

\bibitem{Fu2022} 
C. Fu and C. Chen, 
\href{https://doi.org/10.1088/1475-7516/2023/05/005}
{J. Cosmol. Astropart. Phys. {05} (2023) 005}.

\bibitem{Chen2022}
L. Chen, H. Yu and P. Wu, 
\href{https://doi.org/10.1103/PhysRevD.106.063537}
{Phys. Rev. D \textbf{106}, 063537 (2022)}.

\bibitem{Gu2023}
B. Gu, F. Shu, K. Yang and Y. Zhang, 
\href{https://doi.org/10.1103/PhysRevD.107.023519}
{Phys. Rev. D \textbf{107}, 023519 (2023)}.
	
\bibitem{Yi2022}
 Z. Yi,
 \href{https://doi.org/10.1088/1475-7516/2023/03/048}
 {J. Cosmol. Astropart. Phys. {03} (2023) 048}.

\bibitem{S.Choudhury2023}
S. Choudhury, S. Panda, M. Sami,
\href{https://doi.org/10.48550/arXiv.2303.06066}
{arXiv: 2303.06066}.

\bibitem{Dalianis2023}
I. Dalianis,
\href{https://doi.org/10.48550/arXiv.2310.11581}
{arXiv: 2310.11581}.

\bibitem{Zhao2023}
J-X. Zhao, X-H. Liu, and N. Li
\href{https://doi.org/10.1103/PhysRevD.107.0.43515}
{Phys. Rev. D \textbf{107}, 043515 (2023)}.


\bibitem{Choudhury2023a} 
S. Choudhury, K. Dey, and A. Karde,
\href{https://doi.org/10.48550/arXiv.2311.15065}
{arXiv: 2311.15065}.

\bibitem{Heydari2023} 
S. Heydari, K. Karami,
\href{https://doi.org/10.48550/arXiv.2310.11030}
{arXiv: 2310.11030}.


\bibitem{Firouzjahi2023} 
H. Firouzjahi, A. Talebian,
\href{https://doi.org/10.48550/arXiv.2307.03164}
{arXiv: 2307.03164}.


\bibitem{Mishra2023} 
S. S. Mishra, E. J. Copeland, and A. M. Green,
\href{https://doi.org/10.48550/arXiv.2303.17375}
{arXiv: 2303.17375}.



\bibitem{Asadi2023} 
K. Asadi, A. Nassiri-Rad, and H. Firouzjahi,
\href{https://doi.org/10.48550/arXiv.2304.00577}
{arXiv: 2304.00577}.


\bibitem{Cole2023} 
P. S. Cole, A. D. Gow, C. T. Byrnes, and S. P. Patil,
\href{https://doi.org/10.1088/1475-7516/2023/08/031}
{J. Cosmol. Astropart. Phys. 08 (2023) 031 }.




\bibitem{Choudhury2023b} 
S. Choudhury, S. Panda, and M. Sami,
\href{https://doi.org/10.1088/1475-7516/2023/08/078}
{J. Cosmol. Astropart. Phys. 08 (2023) 078}.




\bibitem{Frolovsky2023} 
D. Frolovsky, S. V. Ketov,
\href{https://doi.org/10.3390/universe9060294}
{{Universe \textbf{2023}, 9(6), 294}.}

\bibitem{Chen2023a} 
C. Chen, A. Ghoshal, Z. Lalak, Y. Luo, and A. Naskar,
\href{https://doi.org/10.1088/1475-7516/2023/08/041}
{J. Cosmol. Astropart. Phys. 08 (2023) 041}.

\bibitem{Su2023} 
B-Y. Su, N. Li, and L. Feng,
\href{https://doi.org/10.48550/arXiv.2306.05364}
{arXiv: 2306.05364}.

\bibitem{Cable2023} 
A. Cable, A. Wilkins,
\href{https://doi.org/10.48550/arXiv.2306.09232}
{arXiv: 2306.09232}.

\bibitem{Escriva2023} 
A. Escriv\`{a}, V. Atal, and J. Garriga,
\href{https://doi.org/10.1088/1475-7516/2023/10/035}
{J. Cosmol. Astropart. Phys. 10 (2023) 035}.

\bibitem{Choudhury2024} 
S. Choudhury, A. Karde, S. Panda, M. Sami
\href{https://doi.org/10.1088/1475-7516/2024/01/012}
{J. Cosmol. Astropart. Phys. 01 (2024) 012}.

\bibitem{Choudhury2023c} 
S. Choudhury,   S. Panda, M. Sami
\href{https://doi.org/10.1088/1475-7516/2023/08/078}
{J. Cosmol. Astropart. Phys. 08 (2024) 078}.


\bibitem{Choudhury2023e} 
S. Choudhury,   K. Dey, A. Karde, S. Panda, and M. Sami
\href{https://doi.org/10.48550/arXiv.2310.11034}
{arXiv: 2310.11034}.

\bibitem{G. Ballesteros}
G. Ballesteros, J. B. Jim\'enez, and M. Pieroni,
\href{https://doi.org/10.1088/1475-7516/2019/06/016}
{J. Cosmol. Astropart. Phys. {06} (2019) 016}.

\bibitem{Kamenshchik}
A. Y. Kamenshchik, A. Tronconi, T. Vardanyan, and G.
Venturi,
\href{https://doi.org/10.1016/j.physletb.2019.02.036}
{Phys. Lett. B \textbf{791}, 201 (2019)}.


\bibitem{Gorji2022}
M. A. Gorji, H. Mothhashi, and S. Mukohyama,
\href{https://doi.org/10.1088/1475-7516/2022/02/030}
{J. Cosmol. Astropart. Phys. {02} (2022) 030}.

\bibitem{Romano3}
A. E. Romano,
\href{https://doi.org/10.48550/arXiv.2006.07321}
{arXiv: 2006.07321}.

\bibitem{Ballesteros2022}
G. Ballesteros, S. C\'{e}spedes, and L. Santoni,
\href{https://doi.org/10.1007/JHEP01(2022)074}
{J. High Energy Phys. 2022 (2022) 74}.


\bibitem{R. Zhai}
R. Zhai, H. Yu, and P. Wu,
\href{https://doi.org/10.1103/PhysRevD.106.023517}
{Phys. Rev. D \textbf{106}, 023517 (2022)}.

\bibitem{Qiu2022}
T. Qiu, W. Wang, and R. Zheng,
\href{https://doi.org/10.1103/PhysRevD.107.083018}
{Phys. Rev. D \textbf{107}, 083018 (2023)}.

\bibitem{Zhai2023}
R. Zhai, H. Yu, and P. Wu,
\href{https://doi.org/10.1103/PhysRevD.108.043529}
{Phys. Rev. D \textbf{108}, 043529 (2023)}.



\bibitem{Corral2023} 
D. del-Corral, P. Gondolo, K. S. Kumar, and J. Marto,
\href{https://doi.org/10.48550/arXiv.2311.02754}
{arXiv: 2311.02754}.



\bibitem{Kasai2023} 
K. Kasai, M. Kawasaki, N. Kitajima, K. Murai, S. Neda, and F. Takahashi,
\href{https://doi.org/10.48550/arXiv.2310.13333}
{arXiv: 2310.13333}.






\bibitem{Stamou2023} 
I. Stamou, S. Clesse,
\href{https://doi.org/10.48550/arXiv.2310.04174}
{arXiv: 2310.04174}.


\bibitem{Ge2023} 
S. Ge, J. Guo, and J. Liu,
\href{https://doi.org/10.48550/arXiv.2309.01739}
{arXiv: 2309.01739}.




\bibitem{Bhattacharya2023} 
G. Bhattacharya, S. Choudhury, K. Dey, S. Ghosh, A. Karde, N. S. Mishra,
\href{https://doi.org/10.48550/arXiv.2309.00973}
{arXiv: 2309.00973}.



\bibitem{Ijaz2023} 
N. Ijaz, M. Mehmood, and M. Rehman,
\href{https://doi.org/10.48550/arXiv.2308.14908}
{arXiv: 2308.14908}.


\bibitem{Choudhury2023d} 
S. Choudhury, A. Karde, S. Panda, and M. Sami,
\href{https://doi.org/10.48550/arXiv.2308.09273}
{arXiv: 2308.09273}.



\bibitem{Cheung2023} 
K. Cheung, C. J. Ouseph, and P-Y. Tseng,
\href{https://doi.org/10.48550/arXiv.2307.08046}
{arXiv: 2307.08046}.




\bibitem{Mansoori2023} 
S. A. H. Mansoori, F. Felegray, A. Talebian, and M. Sami,
\href{https://doi.org/10.48550/arXiv.2307.06757}
{arXiv: 2307.06757}.






\bibitem{Gow2023} 
A. D. Gow, T. Miranda, and S. Nurmi,
\href{https://doi.org/10.48550/arXiv.2307.03078}
{arXiv: 2307.03078}.






\bibitem{Li2023} 
H-J. Li, Y-Q. Peng, W. Chao, and Y-F. Zhou,
\href{https://doi.org/10.48550/arXiv.2304.00939}
{arXiv: 2304.00939}.



\bibitem{Tada2023} 
Y. Tada, M. Yamada,
\href{https://doi.org/10.1103/PhysRevD.107.123539}
{Phys. Rev. D \textbf{107}, 123539 (2023)}.








\bibitem{Cai2023y} 
Y. Cai, M. Zhu, and Y-S. Piao,
\href{https://doi.org/10.48550/arXiv.2305.10933}
{arXiv: 2305.10933}.


\bibitem{Huang2023a}
H. Huang and Y. Piao, 
\href{https://doi.org/10.48550/arXiv.2312.11982}{arXiv: 2312.11982}.







\bibitem{Ghoshal2023} 
A. Ghoshal, A. Moursy, and Q. Shafi,
\href{https://doi.org/10.48550/arXiv.2306.04002}
{arXiv: 2306.04002}.




\bibitem{Saburov2023} 
S. Saburov, S. V. Ketov,
\href{https://doi.org/10.3390/universe9070323}
{Universe \textbf{2023}, 9(7), 323}.


\bibitem{Tada2023a} 
Y. Tada, M. Yamada,
\href{https://doi.org/10.48550/arXiv.2306.07324}
{arXiv: 2306.07324}.



\bibitem{Huang2023} 
H-L. Huang, Y. Cai, J-Q. Jiang, J. Zhang, and Y-S. Piao,
\href{https://doi.org/10.48550/arXiv.2306.17577}
{arXiv: 2306.17577}.



\bibitem{Arya2023} 
R. Arya, R. K. Jain, and A. K. Mishra,
\href{https://doi.org/10.48550/arXiv.2302.08940}
{arXiv: 2302.08940}.







\bibitem{Cai2023i}
Y-F. Cai, C. Tang, G. Mo, et al., 
\href{htps://doi.org/10.48550/arXiv.2301.09403}
{arXiv: 2301.09403}.


\bibitem{Martin2020a}J. Martin, T. Papanikolaou,  and V. Vennin, \href{https://doi.org/10.1088/1475-7516/2020/01/024}
{J. Cosmol. Astropart. Phys. 01 (2020) 024}.

\bibitem{Martin2020b}J. Martin, T. Papanikolaou,  L. Pinol, and V. Vennin, \href{https://doi.org/10.1088/1475-7516/2020/05/003}
{J. Cosmol. Astropart. Phys. 05 (2020) 003}.



\bibitem{Briaud2023}
V. Briaud, V. Vennin, 
\href{https://doi.org/10.1088/1475-7516/2023/06/029}
{J. Cosmol. Astropart. Phys. 06 (2023) 029}.





\bibitem{yfcai2018}
Y. F. Cai, X. Tong, D. G. Wang and S. F. Yan,
\href{https://doi.org/10.1103/PhysRevLett.121.081306}
{Phys. Rev. Lett. \textbf{121},  081306 (2018)}.

\bibitem{yfcai2019}
Y. F. Cai, C. Chen, X. Tong, D. G. Wang and S. F. Yan,
\href{https://doi.org/10.1103/PhysRevD.100.043518}
{Phys. Rev. D \textbf{100}, 043518 (2019)}.

\bibitem{c.chen2019}
C. Chen and Y. F. Cai,
\href{https://doi.org/10.1088/1475-7516/2019/10/068}
{J. Cosmol. Astropart. Phys. {10} (2019) 068}.

\bibitem{c.chen2020}
C. Chen, X. H. Ma and Y. F. Cai,
\href{https://doi.org/10.1103/PhysRevD.102.063526}
{Phys. Rev. D \textbf{102}  063526 (2020)}.


\bibitem{Addazi2022}
A. Addazi, S. Capozziello, and Q. Gan,
\href{https://doi.org/10.1088/1475-7516/2022/08/051}
{J. Cosmol. Astropart. Phys.  {08} (2022) 051}.

\bibitem{Cai2020}
R. Cai, Z. Guo, J. Liu, L. Liu, and X. Yang,
\href{https://doi.org/10.1088/1475-7516/2020/06/013}
{J. Cosmol. Astropart. Phys.  {06} (2020) 013}.

\bibitem{bLi2023}
B. Li, C. Chen, and B. Wang,
\href{https://doi.org/10.48550/arXiv.2307.03747}
{arXiv: 2307.03747}.

\bibitem{Deffayet2011}C. Deffayet, X. Gao, D. A. Steer, and G. Zahariade,
\href{https://doi.org/10.1103/PhysRevD.84.064039} 
{Phys. Rev. D {\bf 84}, 064039 (2011).}

\bibitem{Kobayashi2011}T. Kobayashi, M. Yamaguchi, and J. Yokoyama, \href{https://doi.org/10.1143/PTP.126.511}{Prog.
Theor. Phys. {\bf 126}, 511 (2011).}


\bibitem{Germani2010}
C. Germani and A. Kehagias, 
\href{https://doi.org/10.1103/PhysRevLett.105.011302}
{Phys. Rev. Lett. \textbf{105}, 011302 (2010)}.



\bibitem{S.Tsujikawa2012}	
S. Tsujikawa,
\href{https://doi.org/10.1103/PhysRevD.85.083518}	
{Phys. Rev. D \textbf{85}, 083518 (2012)}.




\bibitem{A.D.Felice2011}	
A. D. Felice and S.Tsujikawa,
\href{https://doi.org/10.1088/1475-7516/2011/04/029}	
{J. Cosmol. Astropart. Phys. {04} (2011) 029}.



%\bibitem{Kobayashi2011}
%T. Kobayashi, M. Yamaguchi, and J. Yokoyama,		
%\href{https://doi.org/10.1143/PTP.126.511}
%{Prog. Theor. Phys. \textbf{126}, 511 (2011)}.


\bibitem{Starobinsky1980}
A. A. Starobinsky,
\href{https://doi.org/10.1016/0370-2693(80)90670-X}
{Phys. Lett. B \textbf{91}, 99 (1980)}.

\bibitem{Babichev2008}
E. Babichev, V. Mukhanov and A. Vikman,
\href{https://doi.org/10.1088/1126-6708/2008/02/101}
{J. High Energy	Phys. \textbf{02} (2008) 101 }.

\bibitem{Armendariz2001}
C. Armendariz-Picon, V. Mukhanov, and P. J. Steinhardt,  
\href{https://doi.org/10.1103/PhysRevD.63.103510}
{Phys. Rev. D \textbf{63}, 103510 (2001)}.


\bibitem{Mukhanov2006}
V. F. Mukhanov and A. Vikman,
\href{https://doi.org/10.1088/1475-7516/2006/02/004}
{J. Cosmol. Astropart. Phys. {02} (2006) 004}.


\bibitem{Armendariz1999}
C. Armendariz-Picon, V. Mukhanov, and P. J. Steinhardt,
\href{https://doi.org/10.1016/S0370-2693(99)00602-4}
{Phys. Lett. B \textbf{458}, 219 (1999)}. 
 
\bibitem{Babichev2006}
E. Babichev, V. Mukhanov,  and A. Vikman, 
\href{https://doi.org/10.1088/1126-6708/2006/09/061}
{J. High Energy	Phys. 09 (2006) 61}.


\bibitem{Kang2007}
J. U. Kang, V. Vanchurin, and S. Winitzki,
\href{https://doi.org/10.1103/PhysRevD.76.083511}
{Phys. Rev. D \textbf{76}, 083511 (2007)}.

\bibitem{Armendariz2005}
C. Armendariz-Picon and E. A. Lim,
\href{https://doi.org/10.1088/1475-7516/2005/08/007}
{J. Cosmol. Astropart. Phys. {08} (2005) 007}.


\bibitem{D.J.Fixsen}
D. J. Fixsen, E. S. Cheng, J. M. Gales, J. C. Mather, R. A. Shafer, and E. L. Wright,
\href{https://doi.org/10.1086/178173}
{Astrophys. J. \textbf{473}, 576 (1996)}.


\bibitem{K.Inomata2016}
K. Inomata, M. Kawasaki, and Y. Tada,
\href{https://doi.org/10.1103/PhysRevD.94.043527}
{Phys. Rev. D \textbf{94}, 043527 (2016)}.


\bibitem{K.Inomata2019}
K. Inomata and T. Nakama,
\href{https://doi.org/10.1103/PhysRevD.99.043511}
{Phys. Rev. D \textbf{99}, 043511 (2019)}.


\bibitem{Young2014}
S. Young, C. T. Byrnes, and M. Sasaki,
\href{https://doi.org/10.1088/1475-7516/2014/07/045}	
{J. Cosmol. Astropart. Phys. {07} (2014) 045}.


\bibitem{Musco2013}
I. Musco and J. C. Miller,
\href{https://doi.org/10.1088/0264-9381/30/14/145009}	
{Class. Quant. Grav. \textbf{30}, 145009 (2013)}.

\bibitem{Harada2013}
T. Harada, C. M. Yoo, and K. Kohri,
\href{https://doi.org/10.1103/PhysRevD.88.084051}
{Phys. Rev. D \textbf{88}, 084051 (2013)}.




\bibitem{B.J.Carr2010}
B. J. Carr, K. Kohri, Y. Sendouda, and J. Yokoyama,
\href{https://doi.org/10.1103/PhysRevD.81.104019}
{Phys. Rev. D \textbf{81}, 104019 (2010)}.

\bibitem{R.Laha}
R. Laha,
\href{https://doi.org/10.1103/PhysRevLett.123.251101}
{Phys. Rev. Lett. \textbf{123}, 251101 (2019)}.





\bibitem{K.Griest}
K. Griest, A. M. Cieplak, and M. J. Lehner,
\href{https://doi.org/10.1103/PhysRevLett.111.181302}
{Phys. Rev. Lett. \textbf{111}, 181302 (2013)}.



\bibitem{P.Tisserand}
P. Tisserand,  L. Le Guillou, C. Afonso, et al. (EROS-2 Collaboration),
\href{https://doi.org/10.1051/0004-6361:20066017}{Astron. Astrophys. \textbf{469}, 387 (2007)}.


\bibitem{V. Poulin}
V. Poulin, P. D. Serpico, F. Calore, S. Clesse and K. Kohri,
\href{https://doi.org/10.1103/PhysRevD.96.083524}
{Phys. Rev. D \textbf{96}, 083524 (2017)}.

\bibitem{Ananda2007}
K. N. Ananda, C. Clarkson, and D. Wands,
\href{https://doi.org/10.1103/PhysRevD.75.123518}
{Phys. Rev. D \textbf{75}, 123518 (2007)}.


\bibitem{Baumann2007}
D. Baumann, P. J. Steinhardt, K. Takahashi and K. Ichiki, 
\href{https://doi.org/10.1103/PhysRevD.76.084019}
{Phys. Rev. D \textbf{76}, 084019 (2007)}.






\bibitem{Kohri2018}
K. Kohri and T. Terada,
\href{https://doi.org/10.1103/PhysRevD.97.123532}
{Phys. Rev. D \textbf{97}, 123532 (2018)}.





\bibitem{ska}
C. L. Carilli and S. Rawlings,
\href{https://doi.org/10.1016/j.newar.2004.09.001}
{New Astron. Rev. \textbf{48}, (2004) 979}.



\bibitem{epta}
L. Lentati, S. R. Taylor,  C. M. F. Mingarelli, et al.,
\href{https://doi.org/10.1093/mnras/stv1538}
{Mon. Not. R. Astron. Soc. \textbf{453}, 2576 (2015)}.


\bibitem{aligo}
J. Aasi, B. P. Abbott, R.  Abbott,  et al. (The LIGO Scientific Collaboration),
\href{https://doi.org/10.1088/0264-9381/32/7/074001}
{Class. Quant. Grav. \textbf{32},  074001 (2015)}.


%\bibitem{Inomata}
%K. Inomata, M. Braglia, and X. Chen, 
%\href{https://doi.org/10.1088/1475-7516/2023/04/011}
%{J. Cosmol. Astropart. Phys. {04} (2023) 011}.


\bibitem{Kristiano2022}
J. Kristiano, J. Yokoyama,
\href{https://doi.org/10.48550/arXiv.2211.03395}
{arXiv: 2211.03395}.

\bibitem{Inomata2023} 
K. Inomata, M. Braglia, and X. Chen, 
\href{https://doi.org/10.1088/1475-7516/2023/04/011} 
{J. Cosmol. Astropart. Phys. 04 (2023) 011.}








\end{thebibliography}
\end{document}